\documentclass[article,
showpacs,preprintnumbers,
 amsmath,amssymb,
floatfix,
]{revtex4-1}

\usepackage{graphicx}
\usepackage{dcolumn}
\usepackage{bm}
\usepackage{color,ulem}

\makeatletter
\gdef\@ptsize{2} 
\let\@currsize\normalsize 
\makeatother
\usepackage{setspace}
\doublespace

\begin{document}


\title{Determination of the origin of the spin Seebeck effect - bulk vs. interface effects}

\author{Andreas Kehlberger}
\altaffiliation[]{Graduate School Materials Science in Mainz, Staudinger Weg 9, 55128, Germany}

\author{Ren\'{e} R\"oser}
\author{Gerhard Jakob}
\author{Mathias Kl\"aui}
\affiliation{Institute of Physics, University of Mainz, 55099 Mainz, Germany}

\author{Ulrike Ritzmann}
\author{Denise Hinzke}
\author{Ulrich Nowak}
\affiliation{Department of Physics, University of Konstanz, D-78457 Konstanz, Germany}

\author{Mehmet C. Onbasli}
\author{Dong Hun Kim}
\author{Caroline A. Ross}
\affiliation{Department of Materials Science and Engineering, Massachusetts Institute of Technology, Cambridge, Massachusetts 02139, USA}

\author{Matthias B. Jungfleisch}
\author{Burkard Hillebrands}
\affiliation{Fachbereich Physik and Landesforschungszentrum OPTIMAS, Technische Universit\"at Kaiserslautern, Kaiserslautern 67663, Germany}

\date{\today}

\begin{abstract}

The observation of the spin Seebeck effect in insulators has meant a breakthrough for spin caloritronics due to the unique ability to generate pure spin currents by thermal excitations in insulating systems without moving charge carriers. Since the recent first observation, the underlying mechanism and the origin of the observed signals have been discussed highly controversially. Here we present a characteristic dependence of the longitudinal spin Seebeck effect amplitude on the thickness of the insulating ferromagnet (YIG). Our measurements show that the observed behavior cannot be explained by any effects originating from the interface, such as magnetic proximity effects in the spin detector (Pt). Comparison to theoretical calculations of thermal magnonic spin currents yields qualitative agreement for the thickness dependence resulting from the finite effective magnon propagation length so that the origin of the effect can be traced to genuine bulk magnonic spin currents ruling out parasitic interface effects.

\end{abstract}

\pacs{72.20.Pa, 72.25.Mk, 75.30.Ds, 85.80.-b}

\maketitle

\section*{\label{sec: Intro}Introduction}

In the fast evolving field of spin caloritronics\cite{Bauer2012} many interesting discoveries have been made, such as the magneto Seebeck effect\cite{Walter2011}, and the spin Seebeck effect (SSE) in metals\cite{Uchida2008}, and semiconductors\cite{Jaworski2010}. One of the most interesting effects is the SSE in ferromagnetic insulators (FMI)\cite{Uchida2010}, such as yttrium iron garnet (YIG). Even in insulators this effect offers the possibility to generate a pure spin current by just thermal excitation. Hence this spin current excited in an insulating system is not carried by moving charge carriers but by excitations of the magnetization, known as magnons. Common theories explain this magnonic SSE, being due to a difference between the phonon- and magnon temperatures $T_{\rm N}$ and $T_{\rm m}$\cite{Xiao2010,Ohe2011}, while other theories rely on a strong local magnon-phonon\cite{Adachi2010} coupling. Up to now no experimental method has been capable of directly observing this temperature difference of magnons and phonons\cite{Agrawal} so that the origin of the genuine SSE in the transverse configuration is still unclear. Furthermore, in the transverse configuration a thermal gradient is generated in the film plane, while the detection layer is on top of the ferromagnetic film. Due to differences in the thermal conductivity of the substrate and ferromagnetic film and the thickness difference between film and substrate as well as the temperature differences between sample and environment, it is challenging to generate only a pure in-plane thermal gradient without an out-of-plane component\cite{Huang2011}. In conductors this out-of-plane component of the thermal gradient will unavoidably lead to parasitic effects, such as the anomalous Nernst effect, that superimpose with any genuine SSE signals in the transverse geometry. For insulators an alternative geometry provides a better controlled configuration. The so called longitudinal configuration\cite{Uchida2010a} establishes the thermal gradient in the out-of-plane direction across the substrate, the ferromagnetic thin film and the detection layer on top. This opens the possibility to study the SSE independently of the thermal conductivity of the substrate, which does not affect the direction of the thermal gradient. Another key point, which complicates the interpretation of the SSE experiments, is the detection method for the thermally excited spin currents. Most spin caloric experiments rely on the indirect measurement by the inverse spin Hall effect (ISHE)\cite{Saitoh2006} to detect the spin current pumped by the SSE. The measured inverse spin Hall voltage\cite{Kajiwara2010,Xiao2010} is predicted to be:

\begin{equation}
V_{\rm ISHE}  = \Theta_{\rm SHE} \,  l_{\rm N} \, \rho_{\rm N} \big<I_{\rm s}^{\rm pump}\big> \approx \, \Theta_{\rm SHE} \,  l_{\rm N} \, \rho_{\rm N} \,\frac{\gamma \hbar g^{\uparrow \downarrow} k_{\rm B}}{\pi M_{\rm s}V_{a}A}\Big(T_{\rm m}-T_{\rm N} \Big)
\label{eq:mix}.
\end{equation}

Here, $\Theta_{\rm SHE}$ is the spin Hall angle of the spin detector material, $l_{\rm N}$ the length between the voltage contacts, and $\rho_{\rm N}$ the resistance of detection layer. The underlying spin current $\big<I_{\rm s}^{\rm pump}\big>$ itself depends on the gyromagnetic ratio $\gamma$, the saturation magnetization $M_{\rm S}$, the spin mixing conductance $g^{\uparrow \downarrow}$, the Boltzmann constant $k_{\rm B}$, the coherence volume of the magnetic system $V_{\rm a}$, the contact area $A$, and the temperature difference of the magnons of the FMI and the phonons of the normal metal (NM) at the interface $\big(T_{\rm m}-T_{\rm N} \big)$. Unfortunately the spin mixing conductance $g^{\uparrow \downarrow}$ and with that the pumped spin currents $I_{\rm s}^{\rm pump}$ are very sensitive to the interface quality\cite{Jungfleisch2013}. This interplay of the interface and the ISHE makes it necessary to carefully maintain the properties of the interface if one wants to compare different samples.
Furthermore parasitic effect may be caused by the detection layer itself: Most experiments use platinum (Pt) for this layer, due to the high spin Hall angle, making it an efficient spin current detector\cite{Mosendz2010}. Many experiments have shown that the paramagnetic Pt shows a measureable magnetoresistance effect in contact with YIG\cite{Huang2012,Weiler2012}. The YIG/Pt interface has been investigated more closely by X-ray magnetic circular dichroism measurements that reveal possibly a small induced magnetic moment of the Pt\cite{Lu2013}. In combination with a thermal gradient, this proximity effect can cause an additional parasitic thermoelectric effect, the anomalous Nernst effect. For this reason one needs even in insulating ferromagnets to clearly distinguish between such parasitic interface effects and the genuine spin Seebeck effects due to spin currents\cite{Qu}. Other recent measurements\cite{Nakayama,Youssef2013} attribute such magnetoresistance effects to a spin Hall magnetoresistance and observe no proximity effect\cite{Geprags2012}. So given these different contradicting claims there is a clear need to distinguish whether the observed signals originate from a parasitic interface effect or a real bulk spin Seebeck effect.\\

Here we present a detailed study of the relevant length scales of the longitudinal SSE in in YIG/Pt by varying the thickness of the ferromagnetic insulator. The obtained results show an increasing and saturating SSE signal with increasing YIG film thickness. By determining also the dependence of the magnetoresistive effect and the saturation magnetization on the thickness, we can exclude an interface effect as the source of the measured signal. By atomistic spin simulation of the propagation of exchange magnons in temperature gradients, we are able to explain this behavior as being due to a finite effective propagation length of the thermally excited magnons.

\section*{\label{sec:Main}Results}

All Y$_3$Fe$_5$O$_{12}$ samples presented in this paper were grown by pulsed laser deposition with film thicknesses ranging from $20\,$nm to $300\,$nm as shown in table~\ref{tab:samples}. The samples have been sorted into three series, where the interface conditions are identical for samples within one series. Details are given in the Methods section.

\begin{table*}
\caption{\label{tab:samples} Thickness of YIG and Pt, crystalline orientation and number of samples for each series.} 
\begin{ruledtabular}
\begin{tabular}{ccccc}
 Series Number & YIG (nm) & Pt (nm) & Orientation & in-situ etching\\
 \hline
 1 & 70,130,200 & 8.5 & 100 & yes \\
 2 & 20,70,130,200,300 & 8.5 & 100 & yes \\
 3 & 40,80,100,130,150 & 10.3 & 100 & no \\
\end{tabular}
\end{ruledtabular}
\end{table*}

\begin{figure}[htb]
\includegraphics[width=1\textwidth]{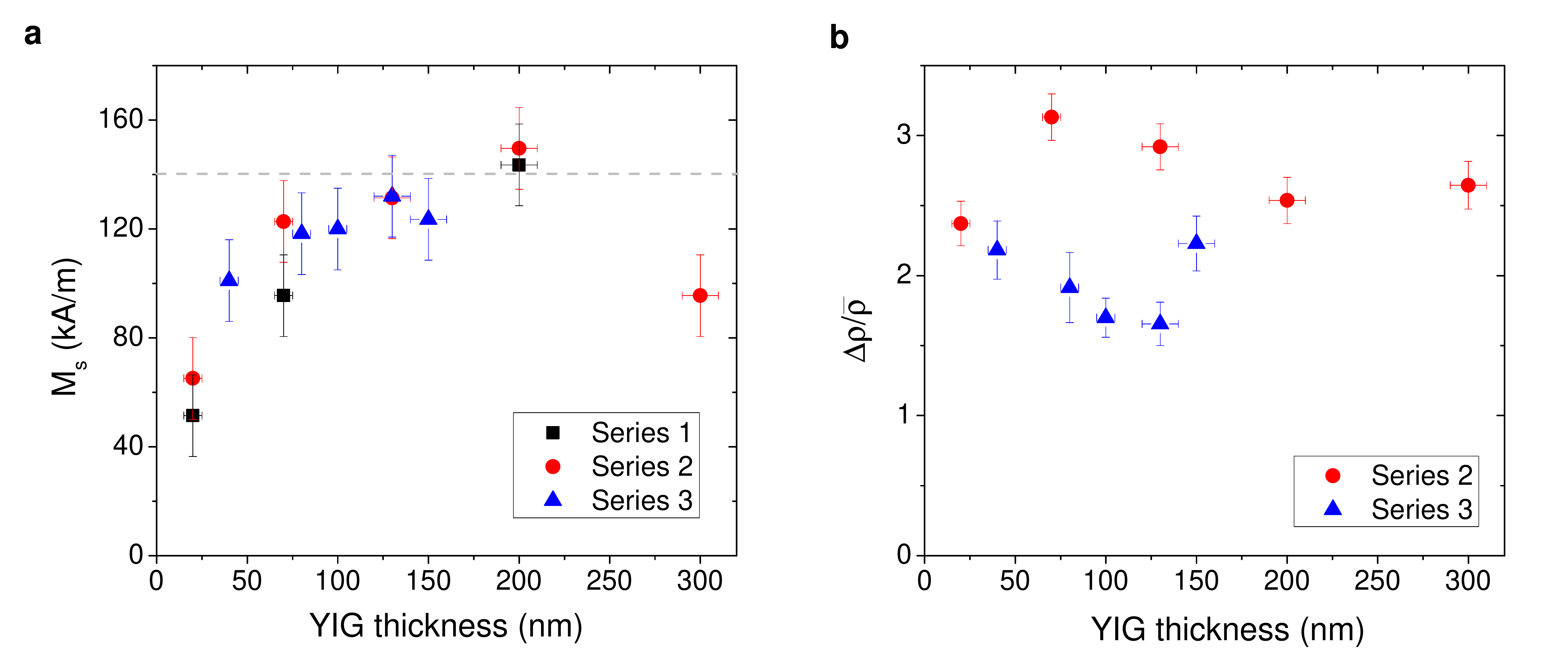}
\caption{\label{fig:MSR} \textbf{Thickness dependence of saturation magnetization and magnetoresistance effect. (a)} Saturation magnetization ($M_{\rm S}$) as a function of YIG film thickness. Each series is marked in different colors. The literature value of $140\,$kA/m is indicated by a grey dashed line. The error of $25$\,kA/m takes into account that the active magnetic volume of the film had to be estimated and the subtraction of the paramagnetic substrate signal. The uncertainty in the thickness determination translates into an error of the active magnetic volume and therefore an error of $M_{\rm S}$.\textbf{(b)} $\Delta \rho / \overline{\rho}$ as a function of YIG-layer thickness measured for series 2 and 3. The y-axis error represents one standard deviation combined with a systematic error considering the temperature variability of the measurement.}
\end{figure}

We first determined the intrinsic magnetic properties of every sample by SQUID magnetometry. Fig.~\ref{fig:MSR}a shows the saturation magnetization ($M_{\rm S}$) as a function of film thickness. Apart from very thin films (20 nm), we find values of approximately $120\,$kA/m$\, \pm 25$\,kA/m, which is close to the literature value of $140\,$kA/m for YIG thin film samples\cite{Serga2010}. For the very thin films, a decrease of the moment has been previously observed for other thin YIG films produced by PLD\cite{Kumara2004}.

To estimate the influence of the YIG/Pt interface coupling, which was previously claimed to be the origin of the measured SSE signals\cite{Huang2012}, we checked the magnetic field dependence of the Pt resistivity. The magnetoresistive effect, which was observable in every sample, showed a dependence on the magnetization direction as well as an in-plane angular dependence that can be explained by the novel spin Hall magnetoresistance effect\cite{Youssef2013,Nakayama}. To determine the correlation with YIG film thickness, we measured the in-plane resistivity for $\theta = 90 ^\circ$ and $\theta = 0 ^\circ$ in a four-point contact configuration. From this data we calculated $\Delta \rho / \overline{\rho} = 2 \left( \rho_{0^\circ} - \rho_{90^\circ}\right) / \left( \rho_{0^\circ} + \rho_{90^\circ}\right)$, as shown in Fig.~\ref{fig:MSR}b. For each series $\Delta \rho / \overline{\rho}$ remained constant, and largely independent of the YIG film thickness. Due to the identical interface conditions for samples of one series, we can assume that the magnetoresistive effect amplitude exhibits no significant dependence on the YIG-film thickness as expected for an interface effect. The changes of the absolute magnetoresistance signal between the different series can be explained by the change of the Pt-thickness and a residual variation of the interface quality.

\begin{figure}[htb]
\includegraphics[width=0.8\textwidth]{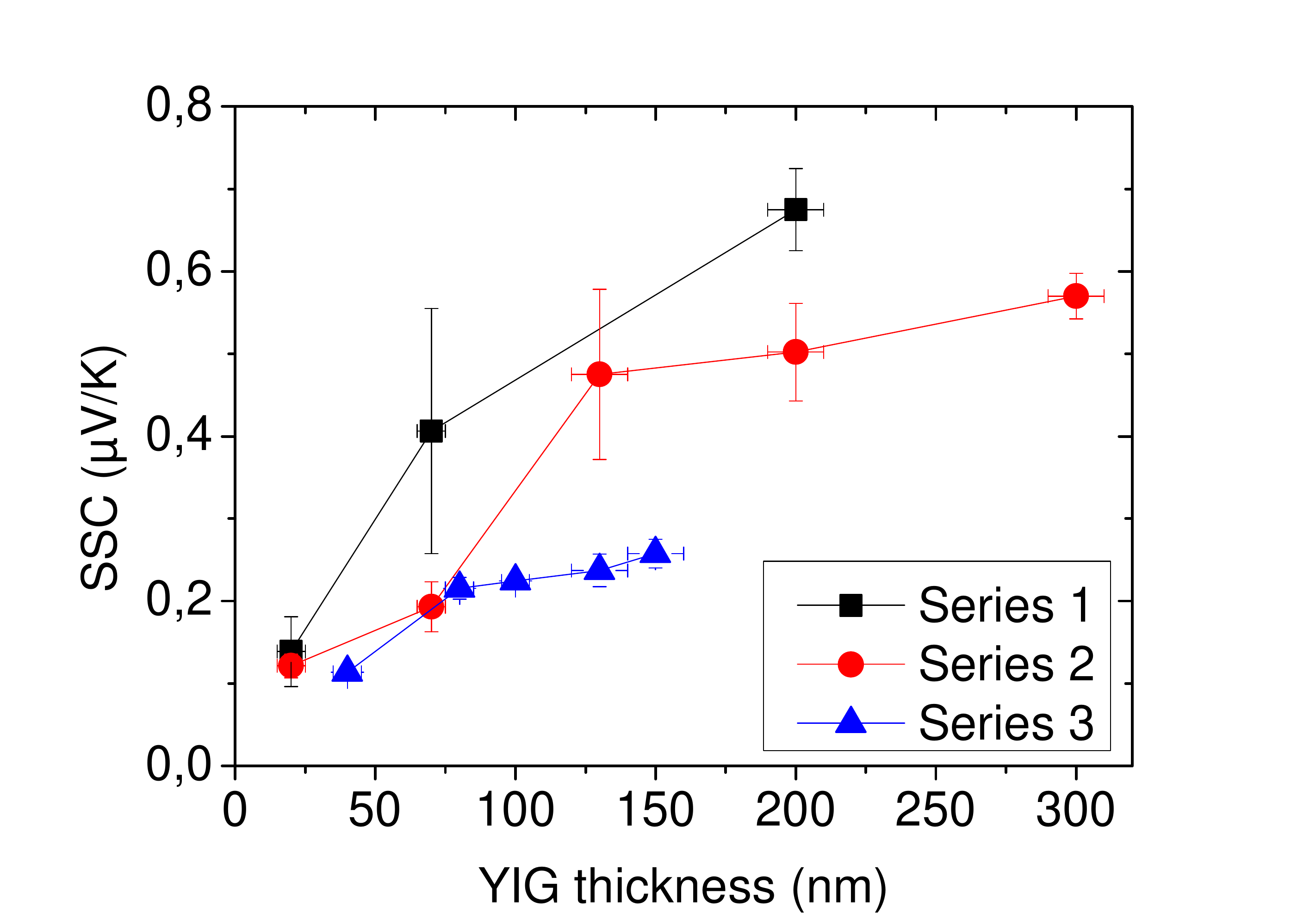}
\caption{\label{fig:SSC} \textbf{Spin Seebeck coefficient as a function of YIG layer thickness}. SSC as a function of YIG-layer thickness. The samples are sorted into different series. Samples of one series have been processed under identical conditions. Data points of each series are connected for clarity. The error in y-axis corresponds to one standard deviation of the measurement data combined with a systematic error taking into account the uncertainty of the mechanical mounting.}
\end{figure}

With the knowledge of the thickness dependence of these material and interface-related parameters, we can now ascertain whether the spin Seebeck effect is correlated to one of those parameters. Three series of YIG films have been investigated in terms of the spin-Seebeck coefficient (SSC), covering a thickness range from $20 \,$nm to $300 \,$nm. A more detailed explanation of the SSC measurements is given in the Methods section. Fig.~\ref{fig:SSC} shows the measured YIG-layer thickness dependence of the SSC for each series.

Below $100\,$nm, films of each series showed an increase of the signal amplitude with increasing thickness. For larger thicknesses the signal starts to saturate. This saturation behavior could be observed in all our series that consist of epitaxial single crystalline films. The samples of series 3 generated signals a factor of two lower than the signals of the other series, due to no in-situ interface etching prior to the Pt deposition, which leads to a less transparent interface for the magnons and therefore a smaller spin mixing conductance\cite{Jungfleisch2013}. This observation underlines the importance of the interface conditions for the comparison of different samples, but the absolute trend of the thickness dependence was not affected by this.

\section*{\label{sec:Discus}Discussion}

In order to understand the origin of the signal, we compare the thickness dependence of the SSC with the thickness dependence of possible underlying mechanisms: When comparing this thickness scaling with that of the saturation magnetization $M_{\rm S}$, shown in Fig.~\ref{fig:MSR}a, we can exclude a direct correlation. We would expect a constant SSC for films thicker than $40\,$nm, since only films below $40\,$nm showed a $M_{\rm S}$ dependence on the YIG thickness. Secondly we compare the thickness dependence of the magnetoresistive effect, shown in Fig.~\ref{fig:MSR}b, with the one of the SSC. Again one would expect a constant contribution to the measured signal independent of the YIG film thickness when comparing with the thickness dependence of the magnetoresistive effect. For this reason we can exclude that any interface coupling effect leads to the observed thickness dependence of the SSC. Even if the magnetoresistive effect in combination with a thermal gradient leads to a Nernst effect, the signals produced by it deliver a constant offset for each series, which cannot be the source of the signal with the thickness dependence that we observe. This is of major importance as it allows us to conclude that the source of the observed signals is not the currently discussed proximity effects at the interface\cite{Huang2012}. The clear thickness dependence points to an origin in the bulk of the YIG.

\begin{figure}[ht]
       \centering
       \includegraphics[width=0.8\textwidth]{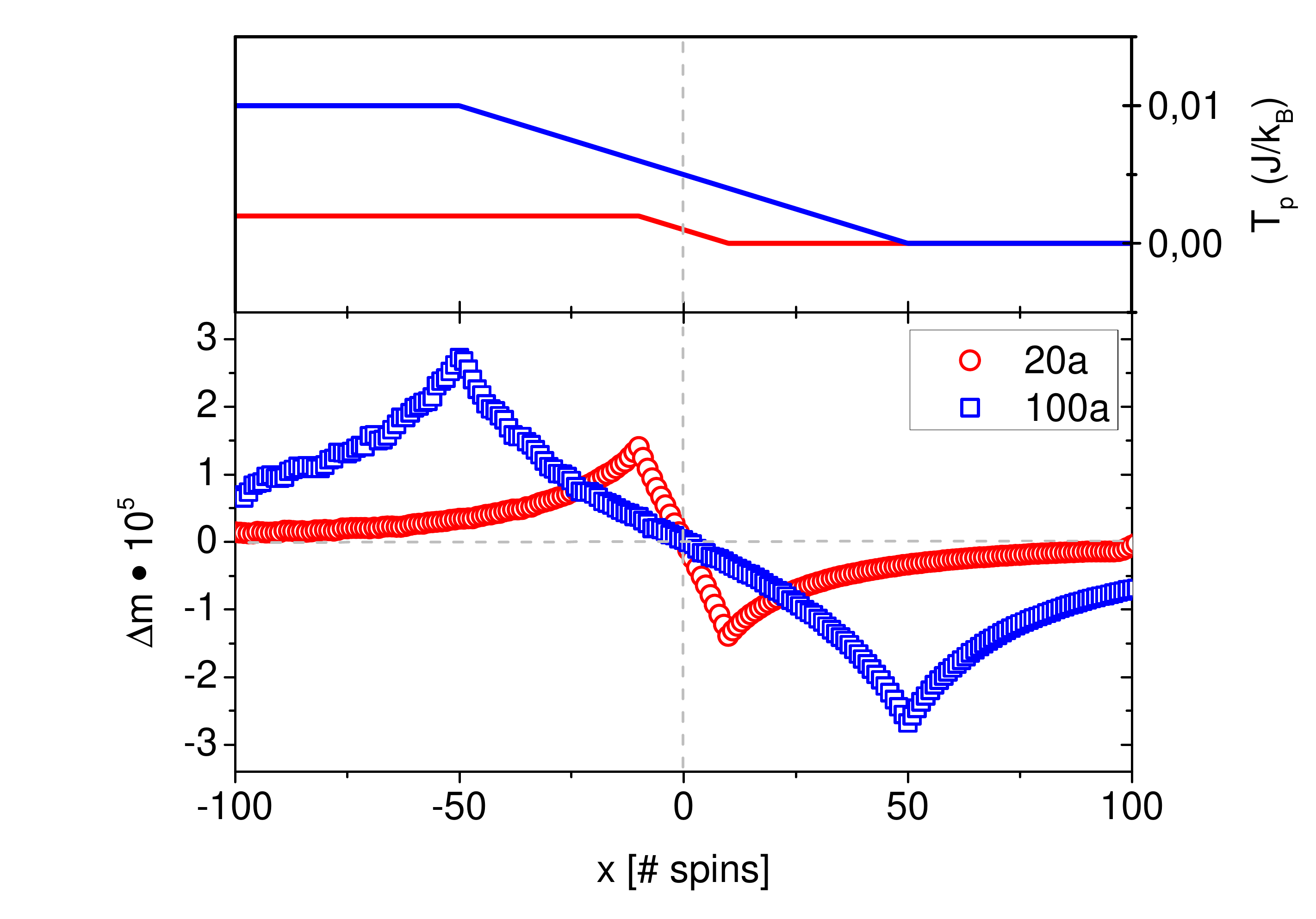}
       \caption{\textbf{Magnon accumulation in a spin system with a temperature gradient.} Magnon accumulation $\Delta m$ as a function of the space coordinate $x$ for a given phonon temperature $T_p$ including a temperature gradient of two different lengths $L=20a,100a$}
       \label{fig1}
    \end{figure}

In the following analysis we assume that the role of the YIG thickness for the SSE might be due to a finite length scale for magnon propagation in the YIG material. In order to investigate this we simulate the propagation of thermally excited magnons in a temperature gradient using an atomistic spin model. The model is generic and not intended to describe YIG quantitatively. It contains a ferromagnetic nearest-neighbor exchange interaction $J$ and an uniaxial anisotropy with easy-axis along $x$-direction and anisotropy constant $d_x=0.1J$. We investigate a cubic system with $512\times8\times8$ spins, which are initialized parallel to the $x$-axis. The dynamics of the spin system is calculated by solving the stochastic Landau-Lifshitz-Gilbert equation numerically with the Heun-Method\cite{ref06}. The phonons provide a heat-bath for the spin system where we assume a linear temperature gradient over the length $L$ in $x$-direction as shown in Fig.~\ref{fig1}. This temperature profile remains  constant during the simulation. After an initial relaxation, the local, reduced magnetization $m(x)$ depends on the space coordinate $x$ and its profile is determined as an average over all spins $\bm{S}_i$ in the corresponding $y$-$z$-plane and additionally as an average over time.

Due to the temperature gradient, magnons propagate from the hotter towards the colder region of the system and this magnonic spin current leads to deviations of the local magnetization $m(x)$ from its local equilibrium value $m_0(x)$ which would follow from the local temperature $T_{\rm p}(x)$ of the phonon system. A temperature dependent calculation of the equilibrium magnetization $m_0(T)$ for a system with constant temperature allows us to describe this deviation, which we define as magnon accumulation $\Delta m(x)$\cite{Ritzmann} via
 
\begin{equation}
 \Delta m(x)=m(x)-m_0(x,T_{\rm p}(x))\;\mbox{.}
	\label{eq:mag}
\end{equation}

Fig.~\ref{fig1} shows this magnon accumulation $\Delta m_x$ as a function of space coordinate $x$ in a system with a damping constant of $\alpha=0.01$ and a temperature profile with a linear temperature gradient of $\Delta T=10^{-5}J/(k_{\rm B}a)$, where $a$ is the lattice constant of the cubic system, for two different lengths $L$ of the temperature gradient. At the hotter end of the gradient magnons propagate towards the cooler region of the system and this reduces the number of the local magnons and increases the local magnetization. On the other side at the cold end of the gradient magnons arriving from hotter parts of the system decrease the local magnetization. The resulting magnon accumulation is symmetric in space and changes its sign in the center of the temperature gradient. The spatial dependence of the magnon accumulation as well as the height of the peaks at the hot and cold end are affected by the mean effective propagation length of the magnons\cite{Ritzmann} in the system. If the length $L$ of the gradient is smaller than the mean propagation length of the magnons, the magnon accumulation depends linearly on the space coordinate $x$. For higher values of the length $L$ the accumulation at the center of the gradient vanishes and appears only at the edges of the temperature gradient. This is in agreement with simulation by Ohe et al. of the transverse spin Seebeck effect\cite{Ohe2011}. In their simulation they modify the mean propagation length of the magnons by changing the damping constant and obtain comparable results.

 \begin{figure}[!ht]
   \centering
   \includegraphics[width=0.8\textwidth]{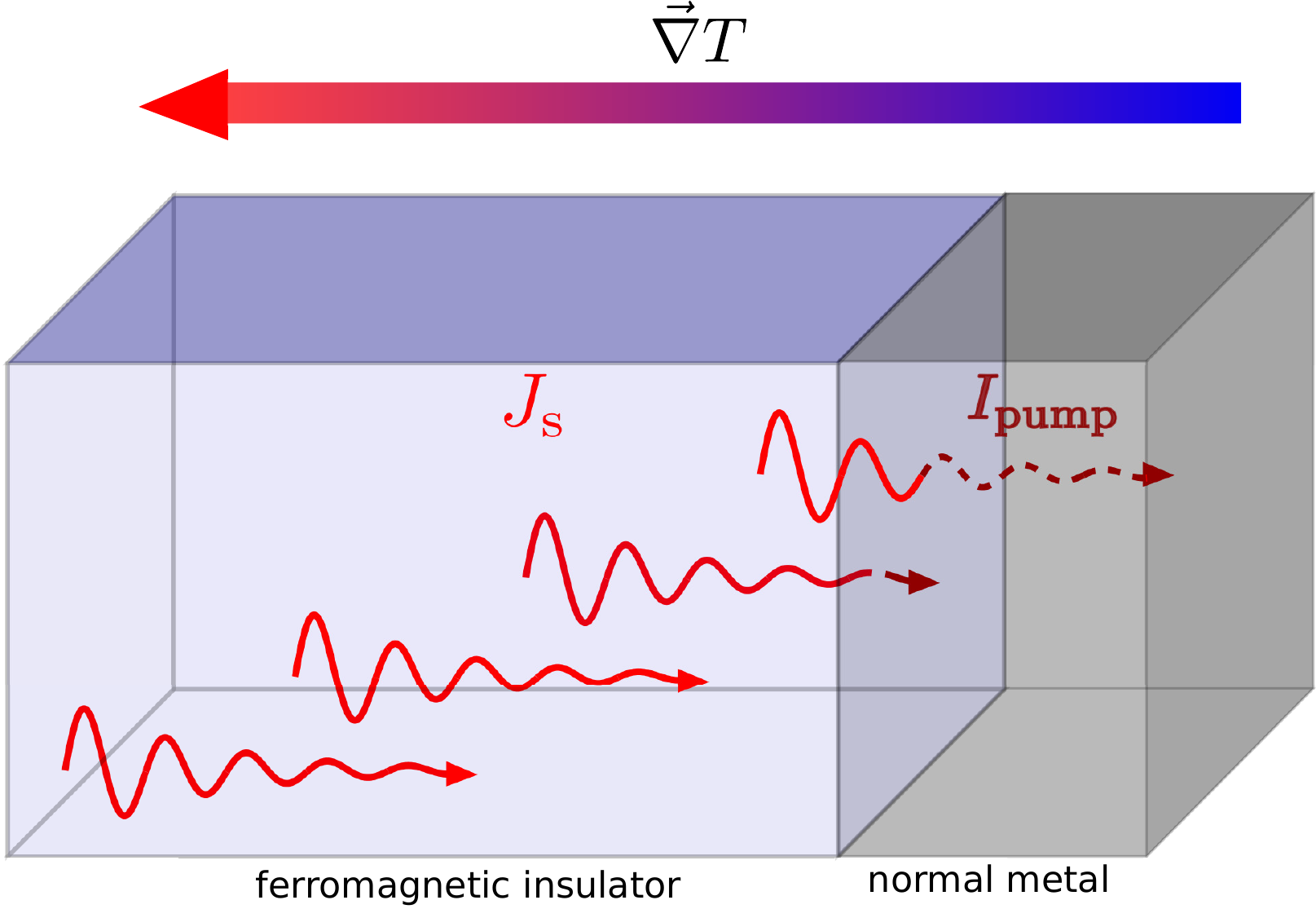}
   \caption{\textbf{Origin of SSC thickness dependence.} Illustration of the saturation effect of the measured voltage due to a finite propagation length of the excited magnonic spin currents.}
   \label{fig3}
 \end{figure}  
The effective mean propagation length $\xi$ of the magnons can be estimated by fitting it to the function.

\begin{figure}[!ht]
   \centering
   \includegraphics[width=1\textwidth]{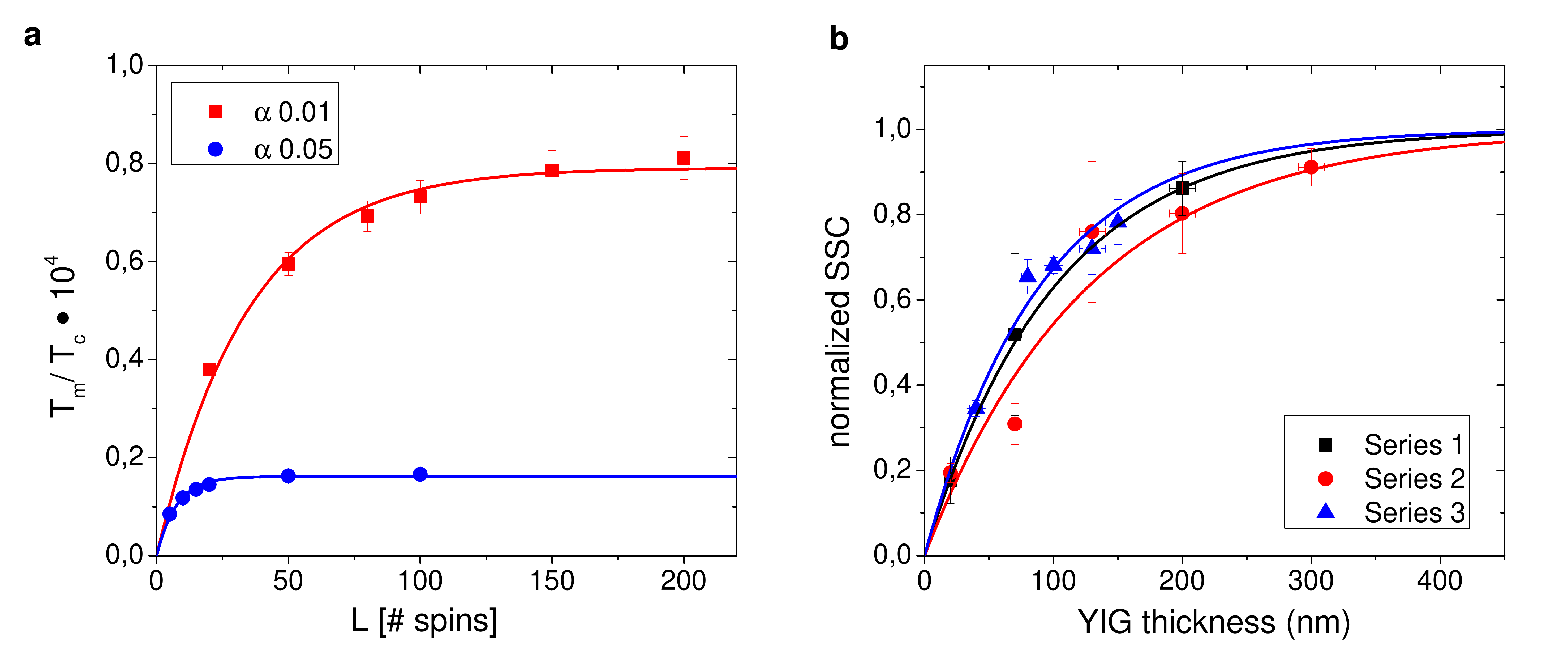}
   \caption{\textbf{Comparison between the theoretical and the experimental results. (a)} Magnon temperature $T_{\rm m}$ at the cold end of the temperature gradient as a function of the length $L$ of the temperature gradient for two different damping constants $\alpha$ shows a saturation effect depending on the propagation length of the thermally excited magnons. \textbf{(b)} Normalized SSC data and corresponding fit functions plotted against the YIG thickness. SSC data have been normalized to the saturation value for an infinitely large system. From the fit we obtain an effective magnon propagation length of $101\,$nm$\pm\,5\,$nm for series 1, $127 \,$nm$\pm\,44\,$nm for series 2 and $89 \,$nm$\pm\,19\,$nm for series 3.}
   \label{fig2}
 \end{figure}

The magnitude of the magnon accumulation at the cold end of the gradient increases with increasing length $L$ up to a saturation value depending on the mean propagation length of the magnons. The magnon accumulation can be understood as the averaged sum of the magnons, which can reach the end of the gradient. As illustrated in Fig. ~\ref{fig3} only those magnons from distances smaller than their propagation length contribute to the resulting magnon accumulation at the cold end of the temperature gradient. Xiao et al. showed that the resulting spin current from the ferromagnet into the non-magnetic material is proportional to the temperature difference between the magnon temperature $T_{\rm m}$ in the ferromagnet and the phonon temperature of the non-magnetic material $T_{\rm N}$\cite{Xiao2010}. Here, for simplicity, we assume that the temperature of the non-magnetic material is $T_{\rm N}=0\;$K and no backflow from the non-magnetic material exists. The magnon temperature $T_{\rm m}$ at the cold end of the gradient can be calculated from the local magnetization $m(x)$. The resulting magnon temperature dependence on the length $L$ of the temperature gradient saturates due to the mean propagation length of the magnons as shown in Fig.~\ref{fig2} for two different damping constants $\alpha$. The variation of the damping constant leads to variation of the mean magnon propagation length and, consequently, different length scales where saturation for the magnon temperature sets in.

\begin{equation}
	 T_{\rm m}^{\rm cold}\propto \Big(1-\exp\Big(-\frac{L}{\xi}\Big) \Big)\mbox{.}
	\label{eq:prop}
\end{equation}

The resulting fits are shown as solid lines in Fig.~\ref{fig2}. The calculated values are comparable to other calculations of the mean propagation length of thermally induced magnons \cite{Ritzmann}. This mean propagation length depends on the frequency spectra of the excited magnons, with that on the model parameters and the damping process during their propagation. The latter depends on the damping constant $\alpha$ as well as the frequency $\omega$ and the group velocity of the magnons $\partial \omega / \partial \bf q\,$ \cite{Kovalev2012}. 

The proportionality between the magnon temperature and the measured ISHE-voltage\cite{Xiao2010} allows us now to evaluate the SSC data points using eq.~\ref{eq:prop}. Each series was evaluated separately as can be seen in Fig.~\ref{fig:SSC}b. We obtain a mean effective magnon propagation length of $101\,$nm$\pm\,5\,$nm for series 1, $127 \,$nm$\pm\,44\,$nm for series 2, and $89 \,$nm$\pm\,19\,$nm for series 3. By this we derive, independent of the different interface qualities between the series, an effective mean propagation length for thermally excited magnons of the order of $110\,$nm$\pm\,16\,$nm from all our series. Based on our model, we can now explain the behavior of the SSC data qualitatively: The increase of the SSC with increasing YIG film thickness below $110\,$nm can be attributed to an increasing magnon accumulation at the interface, while the accumulation starts to saturate and therefore the ISHE-voltage in thicker films. Consequently we can assume that the magnon emitting source is the ferromagnetic thin film and thus we can pinpoint the origin of the observed signal to the magnonic spin Seebeck effect.

In conclusion, we have observed an increasing and saturating spin Seebeck signal with increasing YIG film thickness. This behavior can neither be explained by the thickness dependence of the saturation magnetization nor magnetoresistive effects in the Pt detection layer or any other interface effect. Instead we present a model based on atomistic simulations that attributes this characteristic behavior to a finite propagation length of thermally excited magnons, which are created in the bulk of the ferromagnetic material. From the evaluation of our data we obtain an effective mean propagation length of the order of $110\,$nm for thermally excited magnons, which is in agreement with other studies predicting a finite propagation length of thermally excited magnons of the order of $100\,$nm \cite{Kovalev2012}. Our results thus clearly allow us to rule out parasitic interface effects and identify thermal magnonic spin currents as the source of the observed signals and thus identify unambiguously the longitudinal spin Seebeck effect.

\section*{\label{sec:Method}Methods}

Thin film Y$_3$Fe$_5$O$_{12}$ samples were grown by pulsed laser deposition from a stoichiometric powder target, using a KrF excimer laser ($\lambda$ = $248\,$nm) with a fluence of $2.6\,$J/cm$^2$, and repetition rate of $10\,$Hz\cite{Bi2011}. Monocrystalline $10\,$mm$\times 10\,$mm$\times 0.5\,$mm gadolinium gallium garnet (Gd$_3$Ga$_5$O$_{12}$,GGG) substrates in the (100) crystalline orientation were used to ensure an epitaxial growth of the films, due to the small lattice mismatch below $0.06\,\%$. The optimal deposition conditions were found for a substrate temperature of 650$\,^{\circ}\mathrm{C} \, \pm 30\,^{\circ}\mathrm{C}$ and an oxygen  partial  pressure of $6.67 \times 10^{-3}$ mbar. In order to improve the crystallographic order and to reduce oxygen vacancies, every film was ex-situ annealed at $820\,^{\circ}\mathrm{C}\, \pm 30\,^{\circ}\mathrm{C}$ by rapid temperature annealing for $300\,$s under a steady flow of oxygen. X-ray reflectometry (XRR) and profilometer (Tencor P-16 Surface Profilometer) measurements were done to determine the film thickness, while the crystalline quality was measured by X-ray diffraction.
 
\begin{figure}[htb]
\includegraphics[width=1\textwidth]{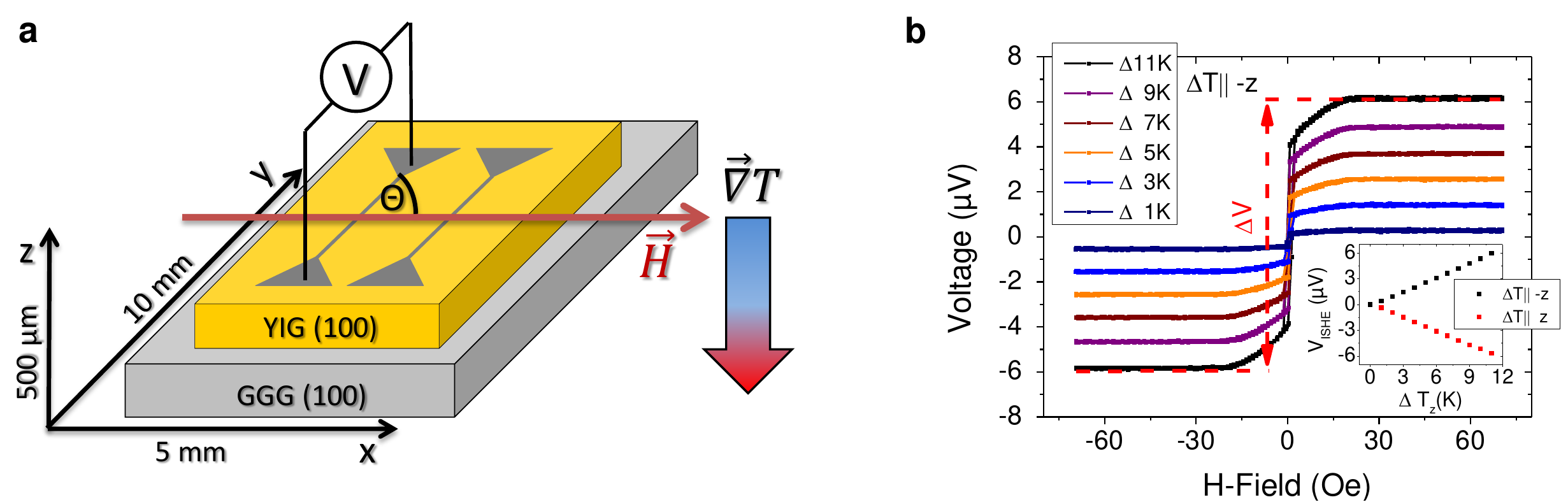}
\caption{\label{fig:Setup} \textbf{Experimental configuration and measured signals. (a)} Sketch of the sample configuration geometry. The structured grey layer, indicates the $4\,$mm long platinum stripes with $~1.2\,$mm large triangular shaped contact pads. The YIG layer is indicated by yellow, the GGG substrate by light grey. For a further understanding of the spin caloric measurements the direction of thermal gradient and the in-plane magnetic field have also been marked. (b): Recorded voltage signals for the $200\,$nm thick YIG film of series 2. Each color represents a different stable temperature difference. The inset shows the evaluated ISHE-voltage $V_{ISHE}$ for both directions of the thermal gradient $\Delta \, T_z$.}
\end{figure}

The samples are sorted into series to highlight the different platinum deposition conditions and therefore interface qualities, which have been used to study interface influence. This influence is minimized for samples within a series by sputtering and cleaning these samples at the same time. Therefore the Pt thickness and the interface preparation were kept identical, while the YIG film thickness varied. Between the series the interface preparations and thus qualities differ and lead to different spin mixing conductance and thus different signal amplitudes for a given thermally excited spin current. For the Pt deposition the samples had been transferred at atmosphere and may therefore have suffered from surface contamination. In order to enhance the interface quality, an in-situ low power ion etching of the YIG surface was performed for some series prior the deposition. DC-magnetron sputtering was used for a homogeneous deposition of the Pt-film under an argon pressure of $1 \times 10^{-2}\,$mbar at room temperature. XRR measurements were done afterwards to control the Pt thickness. In the last fabrication step, the Pt-layer was structured by optical lithography and ion etching in order to reduce influences on the ISHE-voltage by slight variations of the sample geometry. Fig.~\ref{fig:Setup}a shows a sketch of the final sample stack.

For the thermoelectric transport measurements a setup was constructed that is able to generate a temperature difference up to $15\,$K at room temperature in the parallel and anti-parallel out-of-plane direction. Two copper blocks can either serve as heat source or cooling bath to establish a temperature difference between both blocks, while the sample is mounted in between. The relative temperature difference, which was used in the graphs and for the calculation of the spin Seebeck coefficient, was determined by the difference between those two copper blocks. To ensure a good thermal connection, each sample was mounted with thermally conductive adhesive transfer tape. In addition the tape compensated misalignments of the sample mounting. Due to the pressure sensitive heat conduction of the tape, springs were used to mechanically press the two copper blocks together with a constant force in order to reproduce the same conditions for every measurement.\\

Magnons, generated by the thermal gradient in the ferromagnetic layer, will now propagate, depending on the orientation of the thermal gradient, to the FMI/NM interface. An exchange interaction of the local moments of the FMI and the conduction electrons of the NM leads to a spin transfer torque, which creates spin-polarized charge carrier in the NM\cite{Kajiwara2010}. Due to the inverse spin-Hall effect in Pt, a charge carrier separation, based on the spin orientation, is taking place, leading to a measureable potential difference at the edges of the stripe geometry\cite{Saitoh2006, Kajiwara2010}. Our setup used a two point-configuration, shown in Fig.~\ref{fig:Setup}a, to detect this voltage by a nanovoltmeter (Keithley 2182A). By sweeping the in-plane magnetic field with $\theta = 90 ^\circ$ with respect to the platinum stripes in addition to an out-of-plane thermal gradient, one is able to measure magnetic field dependence of the measured voltage as shown as in Fig.~\ref{fig:Setup}b.

To exclude influences of a ground offset, the ISHE-voltage $V_{\rm ISHE}$ needs to be extracted from these voltages signals by dividing the difference of the voltage in saturation for positive and negative H-field of the signals, $\Delta V$, by two. If one now plots the ISHE-Voltage against the corresponding out-of-plane thermal difference $\Delta \, T_z$, as done in the inset of Fig.~\ref{fig:Setup}b the SSC can be derived from the slope by a linear fit. Dependent on the direction of the thermal gradient the SSC will switch its sign, while the absolute value should be the same. The values of the order of $0.54\, \mu V/K$ ($200\,$nm thick YIG film with $8.5\,$nm of Pt) derived for the SSC, in our setup and sample preparation, are similar to experiments of other groups\cite{Uchida2010,Uchida2010a,Weiler2012,Adachi2010,Kirihara2012}. A SSC data point for one sample, as shown in Fig.~\ref{fig:SSC}, is the result of an average of the measurements of the two stripes per sample and for both directions of the thermal gradient.

\bibliography{BibLib_corr}

\subsection*{\label{sec:AuthorCon}Author contributions}

M.K. and G.J. conceived and supervised the research. A.K. and R.R. performed the experiments and analyzed the data. M.C.O., D.H.K.  and C.A.R. provided and characterized the YIG films A.K. and R.R. structured and characterized the samples. U.R., D.H. and U.N. worked on the theoretical modelling. B.J. and B.H. performed additional sample analysis. A.K. organized and wrote the paper with all authors contributing to the discussions and preparation of the manuscript.

\begin{acknowledgments}

The authors would like to thank the Deutsche Forschungsgemeinschaft (DFG) for financial support via SPP 1538 	''Spin Caloric Transport'' and the Graduate School of Excellence Materials Science in Mainz (MAINZ) GSC 266, the EU (IFOX, NMP3-LA-2012246102, MAGWIRE, FP7-ICT-2009-5 257707, MASPIC, ERC-2007-StG 208162) and the National Science Foundation.

\end{acknowledgments}

\end{document}